\documentclass[aps,prd]{revtex4-1}
\usepackage{subfigure,graphics} 
\usepackage{flafter} 
\usepackage{amsmath}
\begin{document} 

\title{
Geometrical terms in the effective Hamiltonian
for rotor molecules } 
\author{Ian G. Moss}
\email{ian.moss@ncl.ac.uk}
\affiliation{School of Mathematics and Statistics, Newcastle University, 
Newcastle Upon Tyne, NE1 7RU, UK}

\date{\today}

%%%%%%%%%%%%%%%%%%%%%%%%%%%%%%%%%%%%%%%%%%%

\begin{abstract}
An analogy between asymmetric rotor molecules and anisotropic cosmology can be used to
calculate new centrifugal distortion terms in the effective potential of asymmetric rotor 
molecules which have no internal 3-fold symmetry. The torsional potential picks up
extra $\cos\alpha$ and $\cos2\alpha$ contributions, which are comparable to
corrections to the momentum terms in methanol and other rotor molecules with isotope
replacements.
\end{abstract}
\pacs{PACS number(s): }
\maketitle

%%%%%%%%%%%%%%%%%%%%%%%%%%%%%%%%%%%%%%%%%%%

Geometrical ideas can often be used to find underlying order in complex systems. In this paper
we shall examine some of the geometry associated with molecular systems
with an internal rotational degree of freedom, and make use of a mathematical analogy 
between these rotor molecules and a class of anisotropic cosmological models to evaluate 
new torsional potential terms in the effective molecular Hamiltonian.

An effective Hamiltonian can be constructed for any dynamical system 
in which the internal forces can be divided into a strong constraining forces and weaker 
non-constraining forces. The surface of constraint
inherits a natural geometry induced by the kinetic energy functional. This geometry
can be described in generalised coordinates $q^a$ by a metric $g_{ab}$ .
The energy levels of the corresponding quantum system divide into energy bands separated by 
energy gaps. A generalisation of the Born-oppenheimer approximation gives an effective Hamiltonian 
for an individual energy band. Working up to order $\hbar^2$, the effective quantum 
Hamiltonian for the reduced theory has a simple form 
\cite{Jensen:1971hc,Maraner:1994nk,Moss:1998jf,Schuster:2003kt}
\begin{equation}
H_{\rm eff}=\frac12|g|^{-1/2}p_a|g|^{1/2}
g^{ab}p_b+V+V_{GBO}
+V_R,\label{kinetic}
\end{equation}
where $p_a=-i\hbar\partial/\partial q^a$, $g^{ab}$ is the inverse of the metric, 
$|g|$ its determinant and $V$ is the 
restriction of the potential of the original system to the constraint surface. 
$V_{GBO}$ depends on forces orthogonal to the constraint surface.
In the lowest energy band, the part depending on the normal mode frequency matrix $\nu$ is
given in Ref. \cite{Moss:1998jf},
\begin{equation}
V_{GBO}=\frac{\hbar}{2}{\rm tr}\,\nu
+\frac{\hbar^2}{16}{\rm tr}\left(\nu^{-1}\nabla\nu\cdot\nu^{-1}\nabla\nu\right),
\label{vgbo}
\end{equation}
where $\nabla_a$ is the Levy-Civita covariant derivative along the constraint surface.
Terms depending on the anharmonicity of the potential can be found in Ref. \cite{Moss:1998jf}.
The next term $V_R$ is purely geometrical in nature
\cite{Jensen:1971hc,Maraner:1994nk},
\begin{equation}
V_R=\frac14 \hbar^2R-\frac18\hbar^2K^2,\label{vr}
\end{equation}
depending on the intrinsic curvature scalar $R$ and the extrinsic curvature scalar $K$ of the
constraint surface. The approximation to the effective Hamiltonian can be extended further
whilst maintaining the symmetry under coordinate redefinitions by including
additional momentum terms, such as
\begin{equation}
\quad \hbar^3\,{\rm tr}(\nu^{-1})R^{ab}\,\nabla_a\nabla_b,
\quad\hbar^3{\rm tr}\left(\nu^{-2}\nabla^a\nu\,\nu^{-1}\nabla^b\nu\right)\nabla_a\nabla_b\dots
\label{extra}
\end{equation}
The underlying geometry implies that the coefficients of these terms are universal.

In molecular systems, constraining forces fix the length of the chemical bonds
leaving the angular orientation of the molecule unconstrained. The potential
$V$ is the sum of the nuclear potential and the usual Born-Openheimer potential, 
$V=V_{NN}+V_{BO}$. Other terms in the effective Hamiltonian are caused by 
rotational-vibrationl coupling.
The commonly adopted procedure is to write down a set of
(non-geometrical) momentum terms and fit their coefficients, centrifugal
distortion constants, using data from molecular spectroscopy 
\cite{watson:1935}. Many of these parameters can then be compared with
their values calculated {\it ab initio} in terms of molecular structure 
constants \cite{watson68}. The geometric expansion of the effective Hamiltonian
(\ref{vgbo}-\ref{extra}) provides a way to repackage this information
in a coordinate independent way.

Rotor molecules have additional internal rotational angles, as shown in figure 1. 
The spectra of rotor molecules are sensitive to the physical environment and to molecular
structure, making these molecules important in many astrophysical and computational
chemistry applications \cite{Kleiner20101}. The aim of the present paper is to investigate the 
geometric potential term $V_R$ in asymmetric rotor molecules. This term vanishes for molecules
with an internal 3-fold symmetry, which are the only rotor molecules in which a formal theory
of the centrifugal distortion has been thouroughly developed 
\cite{Duan1995203,duan:3914,duan:2385,xu:3835,Wang200323}. 
Important aspects of this term are is its universality, and the
fact that it can be calculated relatively simply from the molecular structure. The downside of the
geometrical potential is that it is small compared to leading order terms in the potential which 
can only be evaluated {\it ab initio} by heavy-duty numerical calculations.

\begin{center}
\begin{figure}[htb]
\scalebox{0.4}{\includegraphics{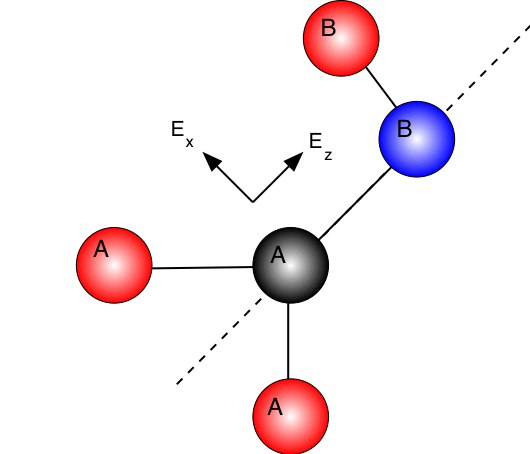}}
\caption{Schematic diagram of a molecule with internal rotation.
Atoms in $A$, the top, can rotate freely with respect to atoms in $B$, the frame.
The dotted line denotes the internal rotation axis,.}
\label{fig1}
\end{figure}
\end{center}

The study of molecules with internal rotation goes back a long way, with many theoretical developments 
made in the 1950's \cite{RevModPhys.31.841}. Some recent methodology can be found in a review by
Kleiner \cite{Kleiner20101}. The basic set-up is shown schematically in figure \ref{fig1}.  
The molecule consists of two sets of atoms $A$ and $B$ 
which are free to rotate independently about a common molecular axis, with angle $\alpha$ 
between $A$ and $B$. The molecule frame is centred on the centre of mass 
with $z$ axis aligned parallel to the internal rotation axis \cite{RevModPhys.31.841} 
and the $x$ axis is fixed with respect to $B$.
The molecule coordinates are the Euler angles $\theta$, $\phi$ and $\psi$ of the
molecular frame and the internal rotation $\alpha$. The approach adopted here is to
develop concepts which are, as far as possible, independent of the choice of coordinates.

The mass the molecule is denoted by $M$ and the mass of $A$ by $M_A$.
A vector $\mbox{\boldmath$\sigma$}$ runs from the centre of mass to a chosen 
point $C$ on the molecular axis so that the locations of the individual nuclei are given 
by ${\bf r}_n=R(\theta,\phi,\psi)(\mbox{\boldmath$\sigma$}+R(0,0,\alpha_n){\bf a}_n)$,
where rotation matrices $R$ act on the vectors ${\bf a}_n$ fixed to $A$ or $B$, and 
$\alpha_n=\alpha$ when $n\in A$, $\alpha_n=0$ otherwise. The total moment
of inertia of the molecule $I_{ij}$ is computed in the molecule frame with origin at the 
centre of mass, whilst the moments of inertia of A and B are more conveniently
centred on the rotation axis at C.

The rotational kinetic energy of the molecule is \cite{quade:540,quade:2512}
\begin{equation}
T=\frac12 I_{ij}\omega^i\omega^j+\rho_i\omega^i\dot\alpha
+\frac12I^A\dot\alpha^2,
\end{equation}
where $\omega^i$ is the angular velocity in the body frame, and the components
are given by 
\begin{eqnarray}
I_{ij}&=&I^A{}_{ij}+I^B{}_{ij}-M({\bf E}_i\times\mbox{\boldmath$\sigma$})\cdot
({\bf E}_j\times\mbox{\boldmath$\sigma$})\\
\rho_i&=&I^A{}_{iz}+M_A({\bf E}_i\times\mbox{\boldmath$\sigma$})\cdot
({\bf E}_z\times\mbox{\boldmath$\rho_A$})\\
I^A&=&I^A{}_{zz}-M_A^2({\bf E}_z\times\mbox{\boldmath$\rho_A$})\cdot
({\bf E}_z\times\mbox{\boldmath$\rho_A$})/M.
\end{eqnarray}
The vector $\mbox{\boldmath$\rho_A$}$ runs from the point C on the rotation axis
to the centre of mass of $A$.

The kinetic energy defines an intrinsic metric $g$ on the four dimensional configuration space. 
It is convenient to use the same notation $\omega^i$ for the differentials associated with 
the angular velocities, and $\omega^\alpha=d\alpha$ for the internal rotation. 
The metric can be written in the form
\begin{equation}
g=I_{ij}(\omega^i+\rho^i\omega^\alpha)(\omega^j+\rho^j\omega^\alpha)
+N^2\omega^{\alpha\,2}.\label{metric}
\end{equation}
Comparison with the kinetic energy shows that the vector $\rho^i=I^{ij}\rho_j$ and 
$N^2=I^A-I^{ij}\rho_i\rho_j$, where $I^{ij}$ is the matrix inverse of $I_{ij}$. 
The metric (\ref{metric}) has a 
three dimensional symmetry group which can be classified under the 
Bianchi classification of Lie algebras as Bianchi type IX \cite{Kundt2003}.
This symmetry corresponds to rotating the molecular coordinate frame and allows us
to change frame to another coordinate system whenever this is convenient.
(This has nothing to do with the molecular symmetry group).
Bianchi IX metrics also arise in the study of cosmological models \cite{ryan1975homogeneous}.
An anisotropic inertia tensor corresponds to an anisotropic cosmological model, and the 
internal angle is analogous to the cosmological time parameter, though a
change from space to spacetime requires replacing $N^2$ by $-N^2$.

The kinetic term in the effective Hamiltonian (\ref{kinetic}) is given by the Laplacian,
\begin{equation}
-\frac{\hbar^2}{2}\nabla^2=\frac12I^{ij}p_ip_j+
|I|^{-1/2}F^{1/2}\left(p_\alpha-\rho^ip_i\right)|I|^{1/2}F^{1/2}
\left(p_\alpha-\rho^jp_j\right),\label{kin}
\end{equation}
where $F=1/2N^2$ and $p_i$ are the usual angular momenta. The momenta can also
be written as $p_a=-i\hbar D_a$, where $D_i$ are the derivatives 
defined by the duality relations $\omega^i(D_j)=\delta^i{}_j$ and 
$D_\alpha=\partial/\partial\alpha$. The ordering of the derivatives in this
way ensures that the quantum theory preserves the Bianchi IX symmetry. Note that whenever the top part part 
of the molecule has $C_{n>2}$ symmetry about the rotation axis, then $F$ and $|I|$ and $\rho^i$ are all 
independent of the internal angle and the factor ordering becomes unimportant. In this case it makes sense to choose
a frame in which selected components of $I_{ij}$ or $\rho^i$ vanish \cite{RevModPhys.31.841}.

The geometrical potential terms in the effective Hamiltonian are invariant under the frame rotations up to
relocation of the origin of the internal rotation angle. The curvature scalar may be calculated using general formulae for 
Bianchi-type metrics in Ryan and Shepley \cite{ryan1975homogeneous}. If ${\bf I}$ denotes the inertia matrix, then
\begin{eqnarray}
R&=&-2F^{1/2}D_\alpha{\rm tr}\left({\bf I}^{-1}F^{1/2}D_\alpha {\bf I}\right)
-\frac12F\left({\rm tr}\,{\bf I}^{-1}D_\alpha {\bf I}\right)^2
-\frac12F\,{\rm tr}\left({\bf I}^{-1}D_\alpha{\bf I}\right)^2\\
&&
+2\,{\rm tr}\,{\bf I}^{-1}+\frac14{\rm tr}\,{\bf I}\,{\rm tr}\,{\bf I}^{-2}-\frac14{\rm tr}\,{\bf I}\,({\rm tr}\,{\bf I}^{-1})^2.
\end{eqnarray}
The remaining extrinsic curvature terms can be related to the tangent vectors ${\bf e}_a=D_a{\bf r}_n$, 
$a=i$ or $\alpha$. 
The components of the intrinsic metric (\ref{metric}) are
\[
g_{ab}=\sum_n m_n(D_a{\bf r}_n)\cdot (D_b{\bf r}_n)
\]
The extrinsic curvature $K_{ab}$ is the normal projection of $D_b{\bf e}_a$,
\begin{equation}
K_{ab}=D_bD_a{\bf r}_n-g^{cd}\sum_p m_p (D_bD_a{\bf r}_p)\cdot D_c{\bf r}_p
\,D_d{\bf r}_n.
\end{equation} 
Its square is given by
\begin{equation}
K^2=g^{ab}g^{cd}\kappa_{abcd}
-g^{ab}g^{cd}g^{ef}\kappa_{abe}\kappa_{cdf}.
\end{equation}
where
\begin{eqnarray}
\kappa_{abcd}&=&\sum_n m_n(D_aD_b{\bf r}_n)\cdot (D_cD_d{\bf r}_n),\\
\kappa_{abc}&=&\sum_n m_n(D_aD_b{\bf r}_n)\cdot D_c{\bf r}_n.
\end{eqnarray}
It is possible to express both of these tensors entirely in terms of the inertia tensors $I_{ij}$, 
$\rho_i$ and $I^A$. If there is a $C_{n>2}$ symmetry about the rotation axis then the 
inertial tensors and the curvature terms are constant. The interesting cases are 
therefore ones in which the molecule has at 
most a $C_2$ symmetry or in which the rotation axis is displaced from the axis of symmetry.

The potential in the effective Hamiltonian has contributions from the zero-point vibrational 
energy and the extra vibrational term in (\ref{vgbo}), which uses the covariant derivatives of 
the normal modes
\begin{equation}
\nabla_i\nu_{IJ}=(\nu_I-\nu_J)\sum_n m_nR(0,0,\alpha_n)\,{\bf d}_{nI}\times{\bf d}_{nJ}.
\end{equation}
${\bf d}_{nI}$ are the displacements of molecule $n$ in normal mode $I$ with frequency $\nu_I$. 
This paper focusses on the less familiar geometrical terms, and the analysis given above
shows that these have no dependence on the vibrational modes.

For the most general type of asymmetric rotor, in an arbitrary molecular frame, the 
geometric potential has a Fourier series expansion
\[
V_R=\sum_{n=0}^\infty \left(a_n\cos\,n\alpha+b_n\sin\,n\alpha\right).
\]
The cofficients can be combined to form a set which is indepenent of the frame, 
by taking $V_0=a_0$ and $V_n=2(a_n^2+b_n^2)^{1/2}.$
These reduce to the usual Fourrier series coefficients when $b_n=0$, which happens, for example,
if a reflection symmetry has been used to align the origin of the molecular frame.

\begin{table}[htb]
\caption{\label{table1}Fourrier series components of the geometric potential term 
$V_R=\frac14\hbar^2 R-\frac18\hbar^2K^2$ and the kinetic function $F$
for a selection of molecules with internal rotation. These use identical molecular structures
for each isotopomer using data obtained from \cite{nist}.
\footnote{A program to calculate these terms can be downloaded from
http://research.ncl.ac.uk/cosmology/publications.html.}}
\begin{ruledtabular}
\renewcommand{\arraystretch}{1.2}
\begin{tabular}{llllllll}
&Isotopomer&$V_1/hc\,(cm^{-1})$&$V_2/hc\,(cm^{-1})$&$V_3/hc\,(cm^{-1})$
&$F_0/hc\,(cm^{-1})$&$F_1/hc\,(cm^{-1})$&$F_2/hc\,(cm^{-1})$\\
\hline
Ethylene
&   ${\rm CH_2CH_2}$&    0.000&    0.279&    0.000&   21.964&    0.000&    0.000\\
&   ${\rm CH_2CD_2}$&    0.000&    0.284&    0.000&   16.476&    0.000&    0.000\\
&   CHDCHD&    0.126&    0.393&    0.003&   15.203&    0.496&    0.048\\
Methanol
&   ${\rm CH_3OH}$&    0.036&    0.001&    0.000&   27.639&    0.129&    0.000\\
&   ${\rm CH_2DOH}$&    0.105&    0.154&    0.000&   26.392&    0.496&    0.060\\
& ${\rm CH_2D{}^{18}OH}$&    0.101&    0.137&    0.000&   26.216&    0.470&    0.051\\
&   ${\rm CHD_2OH}$&    0.040&    0.110&    0.000&   25.480&    0.244&    0.050\\
&   ${\rm CHD_2OD}$&    0.040&    0.126&    0.000&   15.187&    0.244&    0.049\\
Peroxymethyl
&   ${\rm CH_3O_2}$&    0.000&    0.000&    0.000&    6.845&    0.000&    0.000\\
&   ${\rm CH_2DO_2}$&    0.043&    0.291&    0.001&    5.546&    0.359&    0.062\\
&   ${\rm CHD_2O_2}$&    0.015&    0.185&    0.001&    4.684&    0.232&    0.009\\
Nitrosomethane
&   ${\rm CH_3NO}$&    0.017&    0.039&    0.000&    7.680&    0.037&    0.007\\
&   ${\rm CH_2DNO}$&    0.052&    0.297&    0.000&    6.390&    0.412&    0.082\\
&   ${\rm CHD_2NO}$&    0.037&    0.238&    0.001&    5.562&    0.302&    0.047\\
Acetaldehyde
&  ${\rm CH_3CHO}$&    0.011&    0.001&    0.000&    7.744&    0.034&    0.000\\
&  ${\rm CH_2DCHO}$&    0.064&    0.210&    0.001&    6.431&    0.408&    0.059\\
&  ${\rm CHD_2CHO}$&    0.041&    0.140&    0.001&    5.589&    0.301&    0.023
\\
\end{tabular}
\end{ruledtabular}
\end{table}

Results for the geometric potentials $V_R$ and the kinetic term $F$ (see eq. (\ref{kin}))
of a selection of molecules with isotoptic replacements are presented in table 1.
Breaking $C_3$ symmetry results in geometric terms which are typically
in the range 0.1-0.3 $\hbox{cm}^{-1}$. Some of the molecules have non-vanishing potentials 
even though they posses $C_3$ symmetry,  due to misalignment of the rotation axis and the 
axis of symmetry. The components $V_1$ and $F_1$ split doublet into triplet spectral lines in the rotational spectra
of molecules such as ${\rm CD_2HCHO}$ \cite{kilb:1695}. The splitting due to the $F_1$ 
(kinetic) term is known to be insufficient to explain the data without adjusting the alignment
of the rotation axis \cite{quade:540} or introducing empirical centrifugal distortion 
terms \cite{Turner197684}.

With large asymmetry, for example with the halogenated organic molecules 
in table 2, the geometrical potenials can be as high as 1-2 $\hbox{cm}^{-1}$.

\begin{table}[htb]
\caption{\label{table1}Fourrier series components of the geometric potential term 
$V_R=\frac14\hbar^2 R-\frac18\hbar^2K^2$ and the kinetic function $F$
for a selection of halogenated organic molecules with internal rotation.}
\begin{ruledtabular}
\renewcommand{\arraystretch}{1.2}
\begin{tabular}{llllllll}
&Halogen substituent&$V_1/hc\,(cm^{-1})$&$V_2/hc\,(cm^{-1})$&$V_3/hc\,(cm^{-1})$
&$F_0/hc\,(cm^{-1})$&$F_1/hc\,(cm^{-1})$&$F_2/hc\,(cm^{-1})$\\
\hline
Ethylene
&   ${\rm CF_2CH_2}$&    0.000&    0.275&    0.000&   11.555&    0.000&    0.000\\
&   ${\rm CHFCHF}$&    0.594&    1.941&    0.054&    5.291&    2.530&    1.055\\
Methanol
&   ${\rm CHF_2HO}$&    0.284&    0.601&    0.005&   22.353&    0.233&    0.440\\
&   ${\rm CH_2FHO}$&    0.526&    1.090&    0.002&   23.790&    1.318&    0.440\\
&   ${\rm CH_2FFO}$&    0.330&    1.573&    0.059&    4.106&    1.155&    0.746\\
Peroxymethyl
&   ${\rm CH_2FO_2}$&    0.178&    1.378&    0.057&    2.776&    0.899&    0.587\\
&   ${\rm CHF_2O_2}$&    0.145&    0.472&    0.022&    1.665&    0.034&    0.181\\
Nitrosomethane
&   ${\rm CH_2FNO}$&    0.274&    1.542&    0.063&    3.572&    1.084&    0.736\\
&   ${\rm CHF_2NO}$&    0.189&    0.689&    0.038&    2.425&    0.105&    0.398\\
&   ${\rm CF_3NO}$&    0.019&    0.034&    0.000&    2.043&    0.050&    0.024\\
Acetaldehyde
&  ${\rm CH_2FCHO}$&    0.266&    1.128&    0.032&    3.637&    1.120&    0.559\\
&  ${\rm CHF_2CHO}$&    0.135&    0.473&    0.017&    2.485&    0.057&    0.279
\\
\end{tabular}
\end{ruledtabular}
\end{table}

The usefullness, or otherwise, of these geometrical terms is largely dependent on the
accuracy of the calculations for much larger Born-Openhiemer and vibrational potential 
terms. The {\it ab initio} calculations for acetaldehyde in Refs. 
\cite{csaszar:1203,allen:224310} are said to have an accuracy of around 
2 $\hbox{cm}^{-1}$ for the coefficient $V_3$. These calculations include the zero-point vibrational energy 
$\hbar{\rm tr}\,\nu/2$. Note that, in addition to the zero point energy and the geometrical potential 
there should also be the extra vibrational term in the effective Hamiltonian \cite{Moss:1998jf},
\begin{equation}
\frac{\hbar^2}{16}g^{ab}{\rm tr}\left(\nu^{-1}\nabla_a\nu\cdot\nu^{-1}\nabla_b\nu\right).
\end{equation} 

Finally, there are numerical schemes available which go beyond the Born-Openheimer
approximation by quantising the nuclei as well as the electrons 
(e.g. \cite{ishimoto:184309}). These methods give the energy of the ground state
of the molecule, and by comparing different equilibium configurations for the 
internal rotation they can also give an effective barrier height, but they do not 
so far give information about the Fourier components of the effective potential.

%%%%%%%%%%%%%%%%%%%%%%%%%%%%%%%%%%%%%%%%%%%%%
\bibliography{paper.bib}

%merlin.mbs apsrev4-1.bst 2010-07-25 4.21a (PWD, AO, DPC) hacked
%Control: key (0)
%Control: author (8) initials jnrlst
%Control: editor formatted (1) identically to author
%Control: production of article title (-1) disabled
%Control: page (0) single
%Control: year (1) truncated
%Control: production of eprint (0) enabled
\begin{thebibliography}{23}%
\makeatletter
\providecommand \@ifxundefined [1]{%
 \@ifx{#1\undefined}
}%
\providecommand \@ifnum [1]{%
 \ifnum #1\expandafter \@firstoftwo
 \else \expandafter \@secondoftwo
 \fi
}%
\providecommand \@ifx [1]{%
 \ifx #1\expandafter \@firstoftwo
 \else \expandafter \@secondoftwo
 \fi
}%
\providecommand \natexlab [1]{#1}%
\providecommand \enquote  [1]{``#1''}%
\providecommand \bibnamefont  [1]{#1}%
\providecommand \bibfnamefont [1]{#1}%
\providecommand \citenamefont [1]{#1}%
\providecommand \href@noop [0]{\@secondoftwo}%
\providecommand \href [0]{\begingroup \@sanitize@url \@href}%
\providecommand \@href[1]{\@@startlink{#1}\@@href}%
\providecommand \@@href[1]{\endgroup#1\@@endlink}%
\providecommand \@sanitize@url [0]{\catcode `\\12\catcode `\$12\catcode
  `\&12\catcode `\#12\catcode `\^12\catcode `\_12\catcode `\%12\relax}%
\providecommand \@@startlink[1]{}%
\providecommand \@@endlink[0]{}%
\providecommand \url  [0]{\begingroup\@sanitize@url \@url }%
\providecommand \@url [1]{\endgroup\@href {#1}{\urlprefix }}%
\providecommand \urlprefix  [0]{URL }%
\providecommand \Eprint [0]{\href }%
\providecommand \doibase [0]{http://dx.doi.org/}%
\providecommand \selectlanguage [0]{\@gobble}%
\providecommand \bibinfo  [0]{\@secondoftwo}%
\providecommand \bibfield  [0]{\@secondoftwo}%
\providecommand \translation [1]{[#1]}%
\providecommand \BibitemOpen [0]{}%
\providecommand \bibitemStop [0]{}%
\providecommand \bibitemNoStop [0]{.\EOS\space}%
\providecommand \EOS [0]{\spacefactor3000\relax}%
\providecommand \BibitemShut  [1]{\csname bibitem#1\endcsname}%
\let\auto@bib@innerbib\@empty
%</preamble>
\bibitem [{\citenamefont {Jensen}\ and\ \citenamefont
  {Koppe}(1971)}]{Jensen:1971hc}%
  \BibitemOpen
  \bibfield  {author} {\bibinfo {author} {\bibfnamefont {H.}~\bibnamefont
  {Jensen}}\ and\ \bibinfo {author} {\bibfnamefont {H.}~\bibnamefont {Koppe}},\
  }\href {\doibase 10.1016/0003-4916(71)90031-5} {\bibfield  {journal}
  {\bibinfo  {journal} {Annals Phys.}\ }\textbf {\bibinfo {volume} {63}},\
  \bibinfo {pages} {586} (\bibinfo {year} {1971})}\BibitemShut {NoStop}%
%%CITATION = APNYA,63,586;%%
\bibitem [{\citenamefont {Maraner}(1995)}]{Maraner:1994nk}%
  \BibitemOpen
  \bibfield  {author} {\bibinfo {author} {\bibfnamefont {P.}~\bibnamefont
  {Maraner}},\ }\href {\doibase 10.1088/0305-4470/28/10/021} {\bibfield
  {journal} {\bibinfo  {journal} {J.Phys.}\ }\textbf {\bibinfo {volume}
  {A28}},\ \bibinfo {pages} {2939} (\bibinfo {year} {1995})},\ \Eprint
  {http://arxiv.org/abs/hep-th/9409080} {arXiv:hep-th/9409080 [hep-th]}
  \BibitemShut {NoStop}%
%%CITATION = HEP-TH/9409080;%%
\bibitem [{\citenamefont {Moss}\ and\ \citenamefont
  {Shiiki}(2000)}]{Moss:1998jf}%
  \BibitemOpen
  \bibfield  {author} {\bibinfo {author} {\bibfnamefont {I.~G.}\ \bibnamefont
  {Moss}}\ and\ \bibinfo {author} {\bibfnamefont {N.}~\bibnamefont {Shiiki}},\
  }\href {\doibase 10.1016/S0550-3213(99)00650-1} {\bibfield  {journal}
  {\bibinfo  {journal} {Nucl.Phys.}\ }\textbf {\bibinfo {volume} {B565}},\
  \bibinfo {pages} {345} (\bibinfo {year} {2000})},\ \Eprint
  {http://arxiv.org/abs/hep-th/9904023} {arXiv:hep-th/9904023 [hep-th]}
  \BibitemShut {NoStop}%
%%CITATION = HEP-TH/9904023;%%
\bibitem [{\citenamefont {Schuster}\ and\ \citenamefont
  {Jaffe}(2003)}]{Schuster:2003kt}%
  \BibitemOpen
  \bibfield  {author} {\bibinfo {author} {\bibfnamefont {P.}~\bibnamefont
  {Schuster}}\ and\ \bibinfo {author} {\bibfnamefont {R.}~\bibnamefont
  {Jaffe}},\ }\href {\doibase 10.1016/S0003-4916(03)00080-0} {\bibfield
  {journal} {\bibinfo  {journal} {Annals Phys.}\ }\textbf {\bibinfo {volume}
  {307}},\ \bibinfo {pages} {132} (\bibinfo {year} {2003})},\ \Eprint
  {http://arxiv.org/abs/hep-th/0302216} {arXiv:hep-th/0302216 [hep-th]}
  \BibitemShut {NoStop}%
%%CITATION = HEP-TH/0302216;%%
\bibitem [{\citenamefont {Watson}(1967)}]{watson:1935}%
  \BibitemOpen
  \bibfield  {author} {\bibinfo {author} {\bibfnamefont {J.~K.~G.}\
  \bibnamefont {Watson}},\ }\href {\doibase 10.1063/1.1840957} {\bibfield
  {journal} {\bibinfo  {journal} {The Journal of Chemical Physics}\ }\textbf
  {\bibinfo {volume} {46}},\ \bibinfo {pages} {1935} (\bibinfo {year}
  {1967})}\BibitemShut {NoStop}%
\bibitem [{\citenamefont {Watson}(1968)}]{watson68}%
  \BibitemOpen
  \bibfield  {author} {\bibinfo {author} {\bibfnamefont {J.~K.}\ \bibnamefont
  {Watson}},\ }\href {\doibase 10.1080/00268976800101381} {\bibfield  {journal}
  {\bibinfo  {journal} {Molecular Physics}\ }\textbf {\bibinfo {volume} {15}},\
  \bibinfo {pages} {479} (\bibinfo {year} {1968})},\ \Eprint
  {http://arxiv.org/abs/http://www.tandfonline.com/doi/pdf/10.1080/00268976800101381}
  {http://www.tandfonline.com/doi/pdf/10.1080/00268976800101381} \BibitemShut
  {NoStop}%
\bibitem [{\citenamefont {Kleiner}(2010)}]{Kleiner20101}%
  \BibitemOpen
  \bibfield  {author} {\bibinfo {author} {\bibfnamefont {I.}~\bibnamefont
  {Kleiner}},\ }\href {\doibase 10.1016/j.jms.2009.12.011} {\bibfield
  {journal} {\bibinfo  {journal} {Journal of Molecular Spectroscopy}\ }\textbf
  {\bibinfo {volume} {260}},\ \bibinfo {pages} {1 } (\bibinfo {year}
  {2010})}\BibitemShut {NoStop}%
\bibitem [{\citenamefont {Duan}\ and\ \citenamefont
  {Takagi}(1995)}]{Duan1995203}%
  \BibitemOpen
  \bibfield  {author} {\bibinfo {author} {\bibfnamefont {Y.-B.}\ \bibnamefont
  {Duan}}\ and\ \bibinfo {author} {\bibfnamefont {K.}~\bibnamefont {Takagi}},\
  }\href {\doibase 10.1016/0375-9601(95)00642-G} {\bibfield  {journal}
  {\bibinfo  {journal} {Physics Letters A}\ }\textbf {\bibinfo {volume}
  {207}},\ \bibinfo {pages} {203 } (\bibinfo {year} {1995})}\BibitemShut
  {NoStop}%
\bibitem [{\citenamefont {Duan}\ \emph {et~al.}(1996)\citenamefont {Duan},
  \citenamefont {Zhang},\ and\ \citenamefont {Takagi}}]{duan:3914}%
  \BibitemOpen
  \bibfield  {author} {\bibinfo {author} {\bibfnamefont {Y.-B.}\ \bibnamefont
  {Duan}}, \bibinfo {author} {\bibfnamefont {H.-M.}\ \bibnamefont {Zhang}}, \
  and\ \bibinfo {author} {\bibfnamefont {K.}~\bibnamefont {Takagi}},\ }\href
  {\doibase 10.1063/1.471248} {\bibfield  {journal} {\bibinfo  {journal} {The
  Journal of Chemical Physics}\ }\textbf {\bibinfo {volume} {104}},\ \bibinfo
  {pages} {3914} (\bibinfo {year} {1996})}\BibitemShut {NoStop}%
\bibitem [{\citenamefont {Duan}\ \emph {et~al.}(1999)\citenamefont {Duan},
  \citenamefont {Wang}, \citenamefont {Wu}, \citenamefont {Mukhopadhyay},\ and\
  \citenamefont {Takagi}}]{duan:2385}%
  \BibitemOpen
  \bibfield  {author} {\bibinfo {author} {\bibfnamefont {Y.-B.}\ \bibnamefont
  {Duan}}, \bibinfo {author} {\bibfnamefont {L.}~\bibnamefont {Wang}}, \bibinfo
  {author} {\bibfnamefont {X.~T.}\ \bibnamefont {Wu}}, \bibinfo {author}
  {\bibfnamefont {I.}~\bibnamefont {Mukhopadhyay}}, \ and\ \bibinfo {author}
  {\bibfnamefont {K.}~\bibnamefont {Takagi}},\ }\href {\doibase
  10.1063/1.479616} {\bibfield  {journal} {\bibinfo  {journal} {The Journal of
  Chemical Physics}\ }\textbf {\bibinfo {volume} {111}},\ \bibinfo {pages}
  {2385} (\bibinfo {year} {1999})}\BibitemShut {NoStop}%
\bibitem [{\citenamefont {Xu}\ \emph {et~al.}(1999)\citenamefont {Xu},
  \citenamefont {Lees},\ and\ \citenamefont {Hougen}}]{xu:3835}%
  \BibitemOpen
  \bibfield  {author} {\bibinfo {author} {\bibfnamefont {L.-H.}\ \bibnamefont
  {Xu}}, \bibinfo {author} {\bibfnamefont {R.~M.}\ \bibnamefont {Lees}}, \ and\
  \bibinfo {author} {\bibfnamefont {J.~T.}\ \bibnamefont {Hougen}},\ }\href
  {\doibase 10.1063/1.478272} {\bibfield  {journal} {\bibinfo  {journal} {The
  Journal of Chemical Physics}\ }\textbf {\bibinfo {volume} {110}},\ \bibinfo
  {pages} {3835} (\bibinfo {year} {1999})}\BibitemShut {NoStop}%
\bibitem [{\citenamefont {Wang}\ \emph {et~al.}(2003)\citenamefont {Wang},
  \citenamefont {Duan}, \citenamefont {Wang}, \citenamefont {Duan},
  \citenamefont {Mukhopadhyay}, \citenamefont {Perry},\ and\ \citenamefont
  {Takagi}}]{Wang200323}%
  \BibitemOpen
  \bibfield  {author} {\bibinfo {author} {\bibfnamefont {L.}~\bibnamefont
  {Wang}}, \bibinfo {author} {\bibfnamefont {Y.-B.}\ \bibnamefont {Duan}},
  \bibinfo {author} {\bibfnamefont {R.}~\bibnamefont {Wang}}, \bibinfo {author}
  {\bibfnamefont {G.}~\bibnamefont {Duan}}, \bibinfo {author} {\bibfnamefont
  {I.}~\bibnamefont {Mukhopadhyay}}, \bibinfo {author} {\bibfnamefont {D.~S.}\
  \bibnamefont {Perry}}, \ and\ \bibinfo {author} {\bibfnamefont
  {K.}~\bibnamefont {Takagi}},\ }\href {\doibase 10.1016/S0301-0104(03)00251-9}
  {\bibfield  {journal} {\bibinfo  {journal} {Chemical Physics}\ }\textbf
  {\bibinfo {volume} {292}},\ \bibinfo {pages} {23 } (\bibinfo {year}
  {2003})}\BibitemShut {NoStop}%
\bibitem [{\citenamefont {Lin}\ and\ \citenamefont
  {Swalen}(1959)}]{RevModPhys.31.841}%
  \BibitemOpen
  \bibfield  {author} {\bibinfo {author} {\bibfnamefont {C.~C.}\ \bibnamefont
  {Lin}}\ and\ \bibinfo {author} {\bibfnamefont {J.~D.}\ \bibnamefont
  {Swalen}},\ }\href {\doibase 10.1103/RevModPhys.31.841} {\bibfield  {journal}
  {\bibinfo  {journal} {Rev. Mod. Phys.}\ }\textbf {\bibinfo {volume} {31}},\
  \bibinfo {pages} {841} (\bibinfo {year} {1959})}\BibitemShut {NoStop}%
\bibitem [{\citenamefont {Quade}\ and\ \citenamefont {Lin}(1963)}]{quade:540}%
  \BibitemOpen
  \bibfield  {author} {\bibinfo {author} {\bibfnamefont {C.~R.}\ \bibnamefont
  {Quade}}\ and\ \bibinfo {author} {\bibfnamefont {C.~C.}\ \bibnamefont
  {Lin}},\ }\href {\doibase 10.1063/1.1733692} {\bibfield  {journal} {\bibinfo
  {journal} {The Journal of Chemical Physics}\ }\textbf {\bibinfo {volume}
  {38}},\ \bibinfo {pages} {540} (\bibinfo {year} {1963})}\BibitemShut
  {NoStop}%
\bibitem [{\citenamefont {Quade}(1966)}]{quade:2512}%
  \BibitemOpen
  \bibfield  {author} {\bibinfo {author} {\bibfnamefont {C.~R.}\ \bibnamefont
  {Quade}},\ }\href {\doibase 10.1063/1.1727073} {\bibfield  {journal}
  {\bibinfo  {journal} {The Journal of Chemical Physics}\ }\textbf {\bibinfo
  {volume} {44}},\ \bibinfo {pages} {2512} (\bibinfo {year}
  {1966})}\BibitemShut {NoStop}%
\bibitem [{\citenamefont {Kundt}(2003)}]{Kundt2003}%
  \BibitemOpen
  \bibfield  {author} {\bibinfo {author} {\bibfnamefont {W.}~\bibnamefont
  {Kundt}},\ }\href {\doibase 10.1023/A:1022334319617} {\bibfield  {journal}
  {\bibinfo  {journal} {General Relativity and Gravitation}\ }\textbf {\bibinfo
  {volume} {35}},\ \bibinfo {pages} {491} (\bibinfo {year} {2003})}\BibitemShut
  {NoStop}%
\bibitem [{\citenamefont {Ryan}\ and\ \citenamefont
  {Shepley}(1975)}]{ryan1975homogeneous}%
  \BibitemOpen
  \bibfield  {author} {\bibinfo {author} {\bibfnamefont {M.}~\bibnamefont
  {Ryan}}\ and\ \bibinfo {author} {\bibfnamefont {L.}~\bibnamefont {Shepley}},\
  }\href {http://books.google.co.uk/books?id=zaVfQgAACAAJ} {\emph {\bibinfo
  {title} {Homogeneous Relativistic Cosmologies}}},\ Princeton Series in
  Physics\ (\bibinfo  {publisher} {Princeton University Press},\ \bibinfo
  {year} {1975})\BibitemShut {NoStop}%
\bibitem [{\citenamefont {Linstrom}\ and\ \citenamefont
  {Mallard}(2013)}]{nist}%
  \BibitemOpen
  \bibfield  {author} {\bibinfo {author} {\bibfnamefont {P.~J.}\ \bibnamefont
  {Linstrom}}\ and\ \bibinfo {author} {\bibfnamefont {W.~G.}\ \bibnamefont
  {Mallard}},\ }\href {http//webbook.nist.gov} {\emph {\bibinfo {title} {NIST
  Standard Reference Database Number 101}}},\ NIST Chemistry WebBook\ (\bibinfo
   {publisher} {National Institute of Standards and Technology, Gaithersburg},\
  \bibinfo {year} {2013})\BibitemShut {NoStop}%
\bibitem [{\citenamefont {Kilb}\ \emph {et~al.}(1957)\citenamefont {Kilb},
  \citenamefont {Lin},\ and\ \citenamefont {E.~B.~Wilson}}]{kilb:1695}%
  \BibitemOpen
  \bibfield  {author} {\bibinfo {author} {\bibfnamefont {R.~W.}\ \bibnamefont
  {Kilb}}, \bibinfo {author} {\bibfnamefont {C.~C.}\ \bibnamefont {Lin}}, \
  and\ \bibinfo {author} {\bibfnamefont {J.}~\bibnamefont {E.~B.~Wilson}},\
  }\href {\doibase 10.1063/1.1743607} {\bibfield  {journal} {\bibinfo
  {journal} {The Journal of Chemical Physics}\ }\textbf {\bibinfo {volume}
  {26}},\ \bibinfo {pages} {1695} (\bibinfo {year} {1957})}\BibitemShut
  {NoStop}%
\bibitem [{\citenamefont {Turner}\ and\ \citenamefont
  {Cox}(1976)}]{Turner197684}%
  \BibitemOpen
  \bibfield  {author} {\bibinfo {author} {\bibfnamefont {P.~H.}\ \bibnamefont
  {Turner}}\ and\ \bibinfo {author} {\bibfnamefont {A.}~\bibnamefont {Cox}},\
  }\href {\doibase 10.1016/0009-2614(76)80556-8} {\bibfield  {journal}
  {\bibinfo  {journal} {Chemical Physics Letters}\ }\textbf {\bibinfo {volume}
  {42}},\ \bibinfo {pages} {84 } (\bibinfo {year} {1976})}\BibitemShut
  {NoStop}%
\bibitem [{\citenamefont {Csaszar}\ \emph {et~al.}(2004)\citenamefont
  {Csaszar}, \citenamefont {Szalay},\ and\ \citenamefont
  {Senent}}]{csaszar:1203}%
  \BibitemOpen
  \bibfield  {author} {\bibinfo {author} {\bibfnamefont {A.~G.}\ \bibnamefont
  {Csaszar}}, \bibinfo {author} {\bibfnamefont {V.}~\bibnamefont {Szalay}}, \
  and\ \bibinfo {author} {\bibfnamefont {M.~L.}\ \bibnamefont {Senent}},\
  }\href {\doibase 10.1063/1.1633260} {\bibfield  {journal} {\bibinfo
  {journal} {The Journal of Chemical Physics}\ }\textbf {\bibinfo {volume}
  {120}},\ \bibinfo {pages} {1203} (\bibinfo {year} {2004})}\BibitemShut
  {NoStop}%
\bibitem [{\citenamefont {Allen}\ \emph {et~al.}(2006)\citenamefont {Allen},
  \citenamefont {Bodi}, \citenamefont {Szalay},\ and\ \citenamefont
  {Csaszar}}]{allen:224310}%
  \BibitemOpen
  \bibfield  {author} {\bibinfo {author} {\bibfnamefont {W.~D.}\ \bibnamefont
  {Allen}}, \bibinfo {author} {\bibfnamefont {A.}~\bibnamefont {Bodi}},
  \bibinfo {author} {\bibfnamefont {V.}~\bibnamefont {Szalay}}, \ and\ \bibinfo
  {author} {\bibfnamefont {A.~G.}\ \bibnamefont {Csaszar}},\ }\href {\doibase
  10.1063/1.2207614} {\bibfield  {journal} {\bibinfo  {journal} {The Journal of
  Chemical Physics}\ }\textbf {\bibinfo {volume} {124}},\ \bibinfo {eid}
  {224310} (\bibinfo {year} {2006})}\BibitemShut {NoStop}%
\bibitem [{\citenamefont {Ishimoto}\ \emph {et~al.}(2008)\citenamefont
  {Ishimoto}, \citenamefont {Ishihara}, \citenamefont {Teramae}, \citenamefont
  {Baba},\ and\ \citenamefont {Nagashima}}]{ishimoto:184309}%
  \BibitemOpen
  \bibfield  {author} {\bibinfo {author} {\bibfnamefont {T.}~\bibnamefont
  {Ishimoto}}, \bibinfo {author} {\bibfnamefont {Y.}~\bibnamefont {Ishihara}},
  \bibinfo {author} {\bibfnamefont {H.}~\bibnamefont {Teramae}}, \bibinfo
  {author} {\bibfnamefont {M.}~\bibnamefont {Baba}}, \ and\ \bibinfo {author}
  {\bibfnamefont {U.}~\bibnamefont {Nagashima}},\ }\href {\doibase
  10.1063/1.2917149} {\bibfield  {journal} {\bibinfo  {journal} {The Journal of
  Chemical Physics}\ }\textbf {\bibinfo {volume} {128}},\ \bibinfo {eid}
  {184309} (\bibinfo {year} {2008})}\BibitemShut {NoStop}%
\end{thebibliography}%

\end{document}